\documentclass[apl, preprint]{revtex4}
\usepackage{graphicx}
\usepackage{setspace}
\usepackage{amsmath}
\usepackage{amssymb}
\usepackage{latexsym}
\begin{document}
\newcommand {\LSMO} { La$_{2 / 3}$Sr$_{1 / 3}$MnO$_{3}$ }
\newcommand{\YBCO}{ YBa$_{2}$Cu$_{3}$O$_{7}$ }
\newcommand{\PBCO} { PrBa$_{2}$Cu$_{3}$O$_{7}$ }
\newcommand{\cuo}{CuO$_2$ }
\newcommand{\sto}{SrTiO$_3$ }
\newcommand{\twotheta}{2$\theta$ }
\newcommand{\onebar}{$[1\overline{1}0]$}
\title{Growth of [110] \LSMO - \YBCO  heterostructures.}
\author{Soumen Mandal}
\author{Saurabh K Bose}
\author{Rajeev Sharma}
\author{R C Budhani}
\email{rcb@iitk.ac.in} \affiliation{Department of Physics, Indian
Institute of Technology Kanpur, Kanpur - 208016, India}
\author{Prahallad Padhan}
\author{Wilfrid Prellier}
\affiliation{Laboratoire CRISMAT, CNRS UMR 6508, ENSICAEN, 6 Bd du
Marehal Juin, F-14050 Caen Cedex, France}

\begin{abstract}
\begin{spacing}{2}
\YBCO - \LSMO heterostructures of [110] orientation are grown to
allow direct injection of spin polarized holes from the \LSMO into
the \cuo superconducting planes. The magnetic response of the
structure at T $<$ T$_{sc}$ shows both diamagnetic and
ferromagnetic moments with [001] direction as magnetic easy axis.
While the superconducting transition temperature (T$_{sc}$) of
these structures is sharp ($\Delta$T$_{sc} \simeq$ 2.5 K), the
critical current density (J$_c$) follows a dependence of the type
$J_c = J{_o}(1-\frac{T}{T_{sc}})^{\frac{3}{2}}$ with highly
suppressed J$_o$ ($\simeq 2 \times 10^4$ A/cm$^2$) indicating
strong pair breaking effects of the ferromagnetic boundary.
\end{spacing}
\end{abstract}

\maketitle

\begin{spacing}{2}
The transport of quasiparticles and paired electrons across
superconductor (SC) - ferromagnet (FM) proximity effect junctions
provides valuable information on the degree of spin polarization
in the ferromagnet, exchange - field - induced inhomogeneous
superconductivity at SC-FM interface, symmetry of the SC order
parameter and a plethora of other effects arising from the
antagonism between superconductivity and ferromagnetism
\cite{Buzdin, Bergeret, Pokrovsky}, some of which are
technologically important \cite{Gu, Eom}. There have been
extensive studies of polarized and unpolarized quasiparticle
injection in conventional s - wave superconductors \cite{Buzdin,
Pokrovsky, Tedrow}. Similar investigations in heterostructures of
ferromagnetic manganites and hole doped cuprates present a rich
field of research due to the d - wave symmetry of the SC - order
parameter and a high degree of spin polarization in manganites.
One of the systems of interest for such studies is
\YBCO-\LSMO-\YBCO [YBCO - LSMO - YBCO] trilayer. While
magnetotransport and magnetic ordering in such manganite - cuprate
heterostructures and superlattices have been studied in detail
\cite{Larkin, Dong,  Pena, Senapati2}, the YBCO in all such
studies has c-axis perpendicular to the plane of the substrate.
Since superconductivity in YBCO lies in the \cuo planes (ab -
plane), the c-axis oriented structure does not allow injection of
quasiparticles along the nodal or fully gapped directions of the
Fermi surface. In order to overcome this difficulty, it is
necessary to grow the YBCO layer with crystallographic orientation
such that the \cuo planes are normal to the substrate. This can be
achieved by growing either [100]/[010] or [110] oriented YBCO
films. While the [100]/[010] and [110] YBCO oriented films on
lattice matched substrates have been deposited successfully
\cite{Marshall, Inam, Covington}, the growth of an FM - SC
heterostructure or superlattice with the \cuo planes normal to the
plane of the substrate is quite non-trivial.

In this work we report the growth of [110] oriented LSMO - YBCO
heterostructures on [110] \sto which had a 500 {\AA} \PBCO [PBCO]
template. The SC heterostructures show inplane magnetic anisotropy
with [001] as the magnetic easy axis and clear diamagnetic and
ferromagnetic contributions to hysteresis at T $ < $ T$_{sc}$.

Thin films of PBCO - LSMO - YBCO were deposited using  pulsed
laser ablation (PLD) technique\cite{Senapati2}. The optimized
growth temperature (T$_{d}$), oxygen partial pressure pO$_{2}$,
laser energy density (E$_{d}$) and growth rate (G$_{r}$) used for
the deposition of the 500 {\AA} thick PBCO template were,
700$^{0}$ C, 0.4 mbar, $\sim$2 J/cm$^2$ and 1.6 {\AA}/sec
respectively. After the deposition of PBCO layer the substrate
temperature was raised to 750$^{0}$ C keeping the pO$_2$ constant.
The growth of 300 {\AA} thick LSMO was carried out at T$_d$ =
750$^{0}$ C, pO$_2$ = 0.4 mbar, E$_{d}\sim$2 J/cm$^2$ and G$_{r}
\simeq$ 0.5 {\AA}/sec. Once the growth of LSMO layer was complete,
a 1000 {\AA} YBCO film was deposited on top of the PBCO - LSMO
bilayer at T$_d$ = 800$^0$ C, pO$_2$ = 0.4 mbar, E$_{d}\sim$2
J/cm$^2$ and G$_{r} \simeq$ 1.6 {\AA}/sec. After completion of
this layer, the deposition chamber was filled with O$_2$ to
atmospheric pressure and then the sample was cooled to room
temperature with a 30 minutes holdup at 500$^0$ C to realize full
oxygenation of the structure. These deposition parameters were
established after taking a series of calibration runs where the
crystal orientation, high T$_{sc}$ of the YBCO and low coercivity
(H$_c$) of the LSMO layer were important factors in deciding the
best condition.

In Fig. 1 we have shown $\theta - 2\theta$ scans of a PBCO - LSMO
- YBCO heterostructure. Two intense doublets located at $2\theta
\simeq$ 32.5$^o$ and 69$^o$ are seen of which the lower angle
component is due to the [110] and [220] reflections of the
substrate. The weaker component of the doublets which appears at
higher \twotheta is identified with the scattering vector of the
heterostructure normal to the plane of the substrate. However, the
observations of these peaks in $\theta$ - \twotheta scattering
geometry does not confirm [110] phase purity as the reflection
from [103] and [013] oriented phases also fall at the same
\twotheta value. Fig. 2 panel a, b and c respectively show the
$\phi$ scans of [117] peak from [110] and [103] phases of YBCO and
[109] $\phi$ scan for [001] phase of YBCO. In Panel (a) we see
four peaks while panel (b) shows eight. This is due to the fact
that the [117] peaks from [110] phase of YBCO and LSMO lie at the
same position while it is not the case when we are probing the
[117] peaks from the [103] oriented phase. The absence of any peak
in Panel (c) rules out [001] phase. The volume percentage of [110]
oriented grains calculated from the recipe of Westerheim et
al\cite{West} comes out to be $\gtrsim$65\% with the remaining
volume is of [103] grains. While the growth is not 100\% [110]
oriented, the remaining [103] grains still permits injection of
spin polarized carriers directly into the \cuo planes as these
planes are oriented at 45$^o$ with respect to the substrate.

In Fig. 3 we have shown M vs H loops for a PBCO - LSMO - YBCO
heterostructure taken at 10 K with field H parallel to [001] and
\onebar{ } directions of the substrate. The square hysteresis loop
in the main figure (and in inset) when H $\parallel$ [001] as
against a slowly saturating loop when H $\parallel$ \onebar{ }
clearly shows that [001] is the magnetic easy axis of the LSMO. We
can also see distinctly the diamagnetic contribution from the SC,
which splits the field increasing and field decreasing arms of the
loop beyond the saturation field. This splitting is because of the
SC-state is confirmed by the absence of the same in the M-H loop
taken at 200 K (inset). A calculation of the critical current
density (J$_c$) from the shift of M- H loop using Bean
model\cite{later} yields J${_c} \simeq 3.8 \times 10^{4}$ A/cm$^2$
at 10K and 350 Gauss field. A remarkable feature of these
hysteresis loops is the near absence of the diamagnetic
contribution when the magnetic field is aligned along the magnetic
hard axis (\onebar) of LSMO. These features can be understood if
we visualize the way screening currents are induced in the SC-film
by the external field. As shown in Fig. 4, when the external field
is parallel to \cuo planes (H $\parallel$ \onebar), the
diamagnetic moment is produced by weak Josephson tunneling
currents across the \cuo planes. However for H $\parallel$ to
[001], the screening currents are confined to each \cuo plane. A
large condensate density in the planes makes these currents strong
and the diamagnetic moment is distinctly visible in the M - H
loop. The hysteresis loop shift due to superconductivity in
La$_{0.7}$Ca$_{0.3}$MnO$_3$ - YBCO - La$_{0.7}$Ca$_{0.3}$MnO$_3$
superlattices have been reported be Pena et al\cite{Pena}. Their
calculation of J$_c$ on the basis of Bean model yield a
suppression of the current by a factor of 20 at 5K. However, the
work of Ref. 9 is on c-axis oriented films, where the suppression
of the J$_c$ due to ferromagnetic proximity effect may be small as
the c-axis coherence length $\xi_c$ of YBCO is only $\simeq$ 3
\AA.

In Fig. 5 we have shown the variation of transport $J_c$
 with temperature for a PBCO - LSMO - YBCO (circle) and PBCO - YBCO
 (square) heterostructure where PBCO is
500 \AA, LSMO is 300 \AA~ and YBCO is 2000 \AA~ thick, the LSMO
layer being absent in the second structure while other thicknesses
are same. The SC - transition in these samples measured in a four
probe geometry is shown in the inset of Fig.5. We note that while
this transition in PBCO - LSMO - YBCO heterostructure is quite
sharp ($\Delta$T$_{sc}\thickapprox 2.5$ K) as compared to the
[110] oriented PBCO - YBCO bilayer, its J$_c$ is greatly
suppressed. The J$_c$(T) data have been fitted to the
phenomenological relation $J_c = J{_o}(1-\frac{T}{T_{sc}})^\beta$.
The fitting parameters for PBCO - LSMO - YBCO and PBCO - YBCO
structures are $J_o = 2.0 \times 10^4$ A/cm$^2$ and $1.8 \times
10^5$ A/cm$^2$ respectively, while $\beta = 1.5$ and 2.16
respectively. In the Ginzburg - Landau description of J$_c$(T) the
prefactor J$_o$ is directly related to the condensate density. A
highly suppressed J$_o$ in samples with ferromagnetic boundary
provides a strong indication of pair breaking by spin polarized
carriers injected from the LSMO.

 In summary, manganite - cuprate bilayers where \cuo
 planes are normal to the plane of the templated
[110] \sto have been synthesized. This structure is amenable to
deposition of a second low - coercivity LSMO film on top of the
YBCO. Studies of the ferromagnetism and superconductivity in such
[110] oriented FM - SC - FM structures are in progress.

We acknowledge financial support from the Indo - French Center for
Promotion of Advanced Research and the Department of Defence,
Government of India.
\end{spacing}

\newpage

\begin{figure}
\centerline{\includegraphics[width=5.88in]{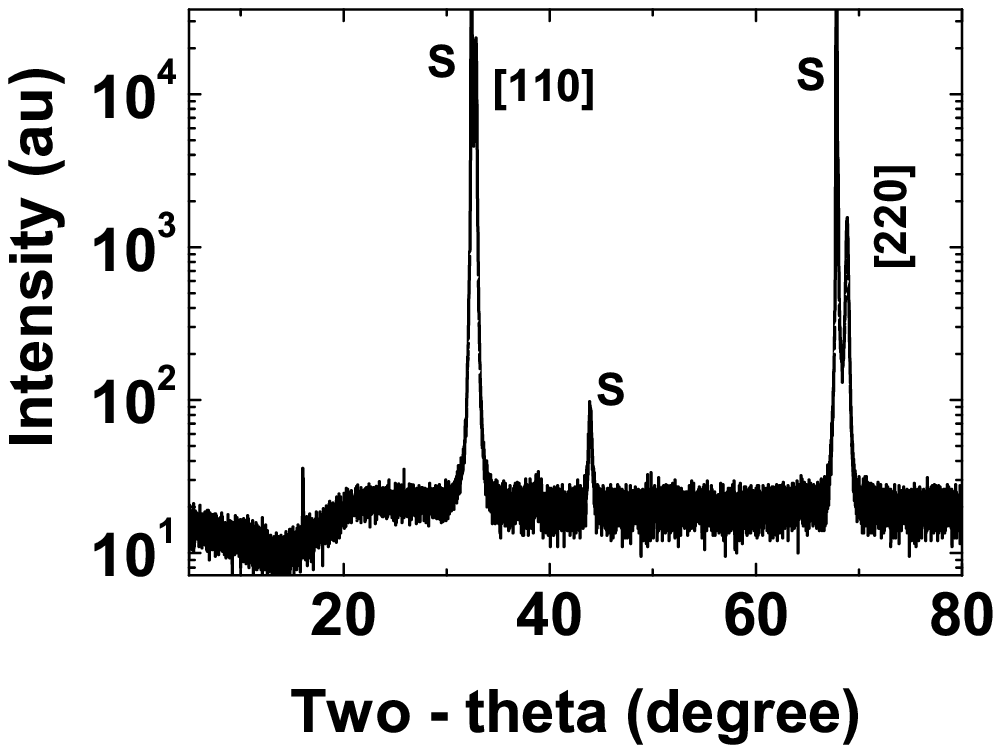}}
\caption{$\theta - 2\theta$ X - ray diffraction pattern of PBCO -
LSMO -
YBCO heterostructure grown on [110] STO.}\label{fig1}
\end{figure}

\newpage

\begin{figure}
\centerline{\includegraphics[width=5.88in]{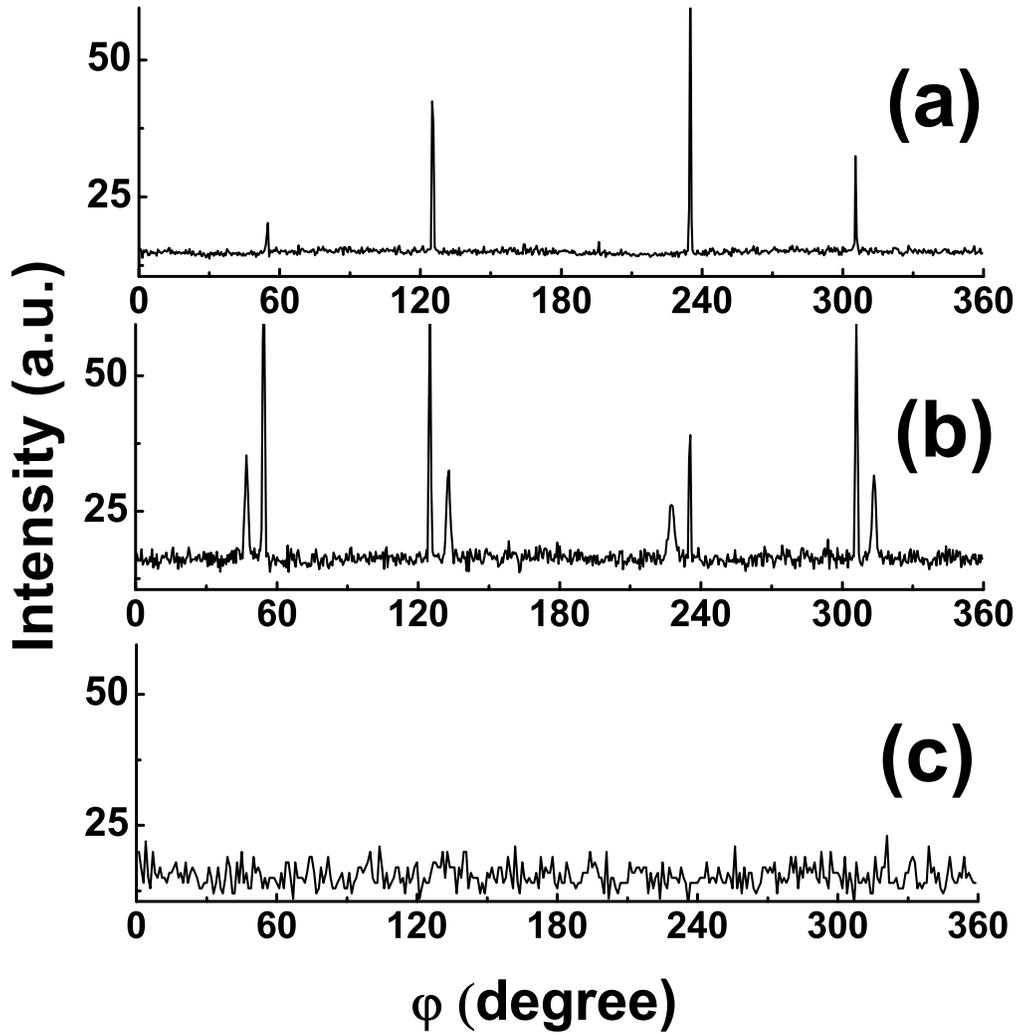}}\caption{$\phi$ scans of the 
PBCO - LSMO - YBCO heterostructure. Panel (a) and (b) show the [117] $\phi$ scan for
[110] and [103] phases respectively, clearly indicating the
presence of both [103]/[110] phases of YBCO. Panel (c) shows the
[109] $\phi$ scan for [001]phase indicating the absence of the same.}\label{fig2}
\end{figure}

\newpage

\begin{figure}
\centerline{\includegraphics[width=5.88in]{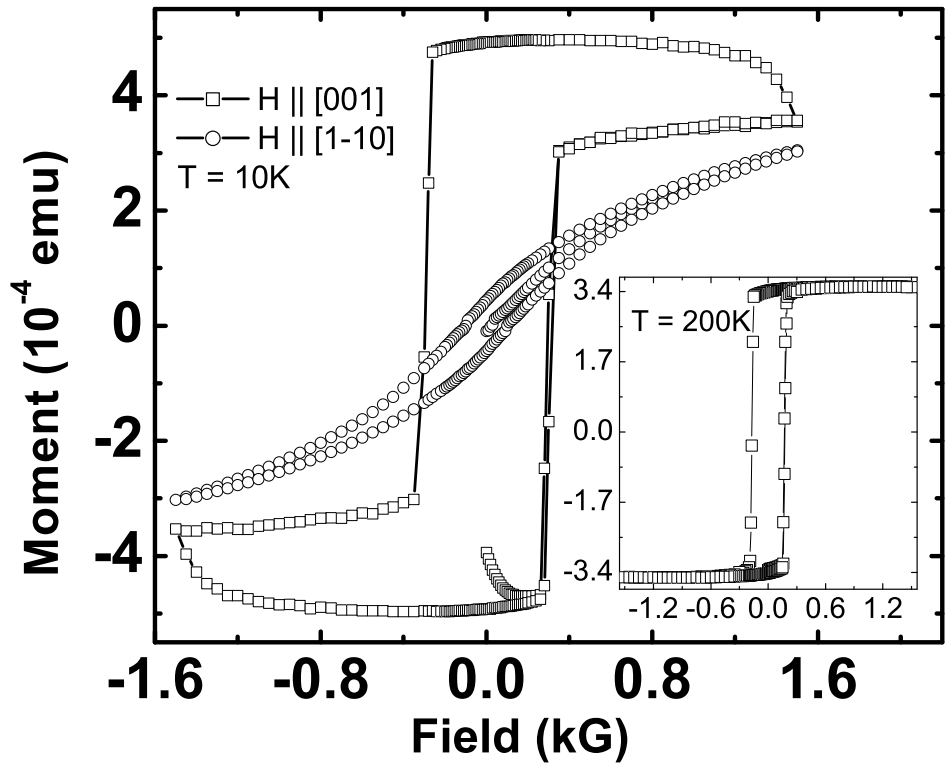}}\caption{M vs H loops for PBCO - LSMO - YBCO
heterostructures taken at 10 K with field $\parallel$ and $\perp$
to \onebar { } substrate edge. The inset shows the M - H loop with
H along the magnetic easy axis at 200 K.}\label{fig3}
\end{figure}

\newpage
\begin{figure}
\centerline{\includegraphics[width=5.88in]{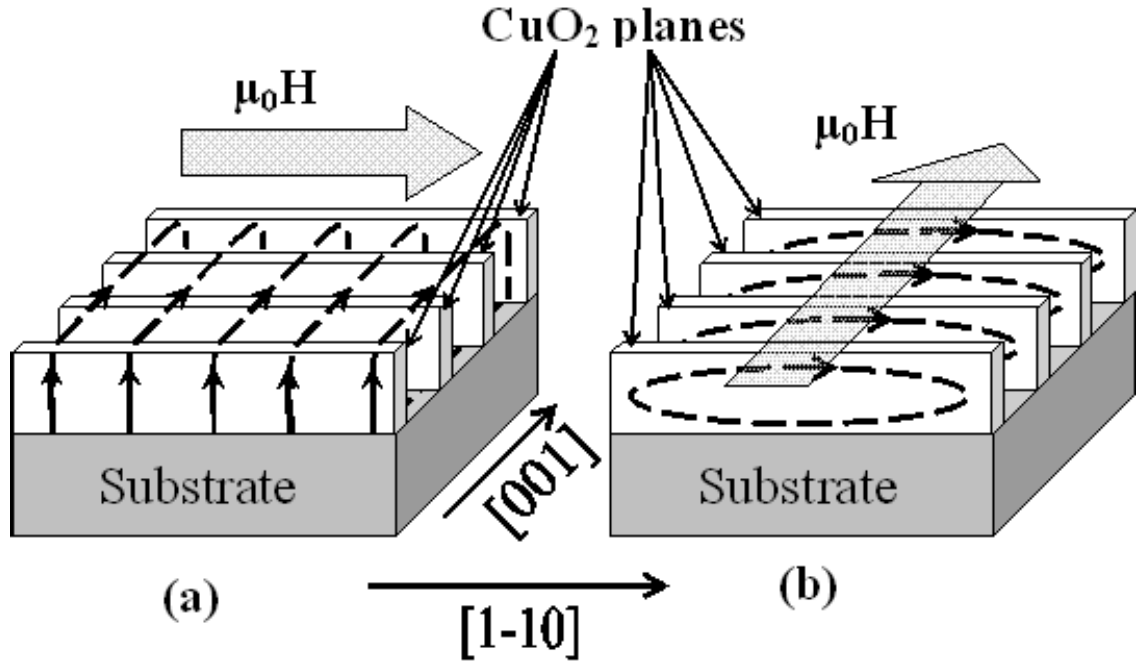}}\caption{a) Josephson 
tunneling controlled screening
current in [110] oriented heterostructure when the applied field H
$\parallel$ \cuo planes. (b) In - plane screening current in [110]
oriented heterostructure when the applied field H $\perp$ \cuo planes.}\label{fig5}
\end{figure}

\newpage
\begin{figure}
\centerline{\includegraphics[width=5.88in]{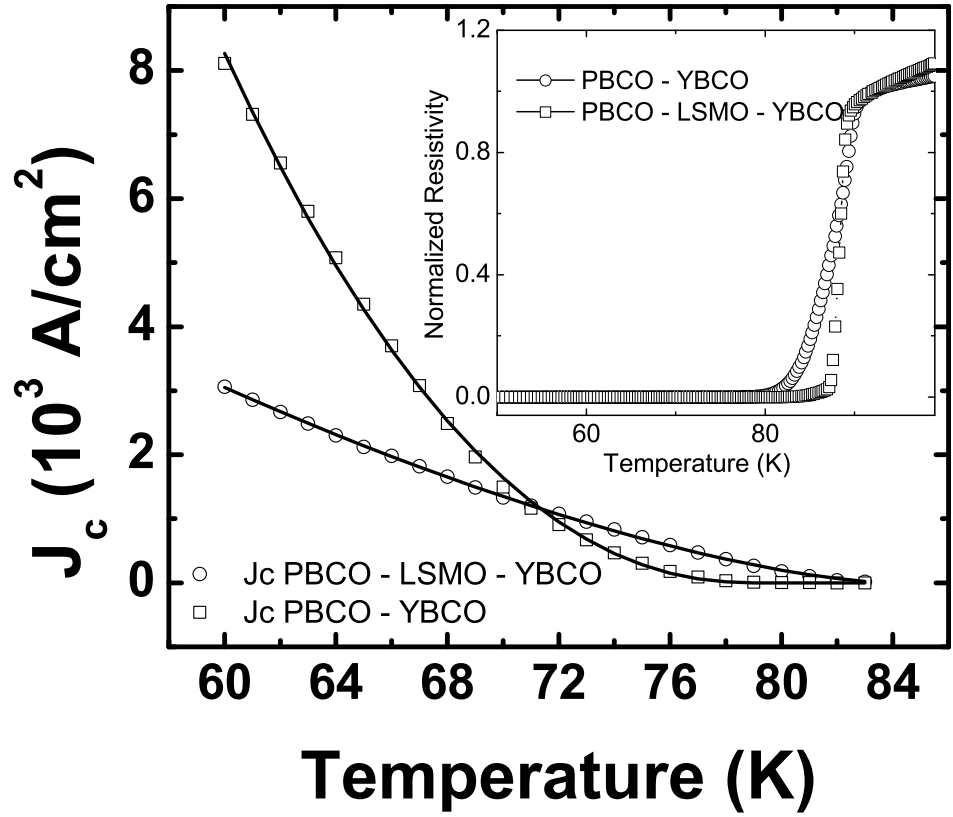}}
\caption{Transport J$_c$(T) data for PBCO - YBCO and PBCO
- LSMO - YBCO heterostructures taken with voltage criterion of 10
$\mu$V/cm. The solid lines are the fitting using the formula $J_c
= J{_o}(1-\frac{T}{T_{sc}})^\beta$. The inset shows the
resistivity of the same samples in the vicinity of the T$_{sc}$.
The resistivity has been normalized with respect to its
value at 92 K.}\label{fig4}
\end{figure}

\end{document}